\documentstyle[12pt,myjcp]{report}

\begin{document}

\noindent

{\bf I. Introduction}

\vskip .1in

Whether you want to plan an efficient bus route, optimize the
positions of integrated circuits on a silicon chip, or predict the
lowest energy configuration of a cluster of atoms, the problem reduces
to one of finding minima or maxima of specified functions.  The
significance of the optimization field is reflected in the number of
papers written on the subject.  Over 300 articles on one method alone,
simulated annealing, have been published since 1988.

There appears to be no one ``perfect'' algorithm that will solve 
every optimization problem.  Instead, a host of complementary
methods have evolved, each being suited to specific tasks [1].  For
example, Brent's method or the golden section search can be used if the
function to be minimized (or maximized) has only one independent variable
even if its derivative is unknown.  Multidimensional functions are much
more difficult to optimize.  If the first derivatives can be
calculated, conjugate gradient methods or variable metric methods may be
used.  Otherwise, the downhill simplex method, direction-set
methods, simulated annealing, or genetic annealing may be used.  None of these
techniques can be universally implemented, and the method of choice
varies with the details of the specific problem.

Simulated annealing is a particularly promising minimization
technique [1,2].  It has, for example, proved effective in finding the
global minimum of multidimensional functions having large numbers
of local minima.  As with other Monte Carlo based approaches, this
method is well suited for implementation on both serial and parallel
architectures.  The mathematical problem, the optimization of a function, is
``solved'' by relating the task to a corresponding physical process,
thermal annealing.  As classical systems are slowly cooled, the
configuration space density tends to condense in regions where the
potential energy is small.  If the cooling process is slow enough, then the
system ultimately finds the physical arrangement that minimizes its
potential energy.  Hence, by assigning the function to be minimized to
be the analog of the potential energy and some control parameter as
the analog of temperature, the global minimum of a function can be
found by simulating the annealing process as the ``temperature'' is
taken to zero.  As in the case of the physical system, the rate of
cooling is important in determining whether or not such annealing
procedures ultimately find their way to the global minimum or become
trapped in various local minima.  Although this technique is relatively
new, it has proved useful for a wide range of optimization problems [1].

Within the protein folding community, various quantum methods have been
developed in which the global minimum is found either by tunneling out
of local minima or by removing uninteresting minima through smoothing
[3].  While this work was in progress, Amara, Hsu, and Straub developed
a minimization scheme for multidimensional potentials via an approximate
solution of the imaginary time Schr\"odinger equation.  The approximate 
wave function is comprised of a Hartree product of single wave packets.
These packets are allowed to move, tunnel, expand, and contract in
search of the global minimum.  The quantum mechanics of the system is
then relaxed.  In the limit of h -$>$ 0, the classical minimum
if found [4].

We describe in Section II a new optimization approach, quantum
annealing, that is closely related to its classical counterpart.  Unlike
the other quantum mechanical approaches discussed, this method does not
require an approximation to the wavefunction.  We illustrate its use with
two examples in Section III.  The first, a pedagogical exercise, is
designed to illustrate how quantum annealing avoids local minima.  The
second is a non-trivial determination of the lowest energy
configurations of various homogeneous Lennard-Jones clusters.  In
Section IV we present our summary and discussion.

\vskip .2in

{\bf II.  Formal Development}

\vskip .2in

We begin by assuming that our task is the minimization of a specified,
many-variable function.  As in conventional annealing methods, it is
convenient to view this function as being the potential energy of a
hypothetical physical system.  in contrast to the simulated annealing
approach, however, it proves convenient to view our system as
quantum-mechanical rather than classical in nature.

Assuming that we can compute the average energy of our system as a
function of its temperatue and quantum mechanical character, we can
approach our objective, the minimum of the system's potential energy
(point A in Fig. (1)), in a variety of ways.  Starting at the arbitrary
point D in Fig. (1), simulated annealing first
turns off the system's quantum mechanics and then reduces the
temperature of the resulting classical system to zero (path DBA in Fig.
(1)).  It proves useful, however, to invert the order of these two
limits first reducing the system's temperature to zero and then
following the resulting quantum ground state energy to its classical
limit (path DCA in Fig. (1)).  Since the fictitious mechanical system
involved is an artificial construct, we are free to craft its character
to suit our purposes.  In particular, we can control the degree of its
quantum mechanical behavior by varying the masses of its constituent
particles.

The motivation for inverting the usual order of the limits is that, in a
sense, it is easier to take the zero temperature limit of a quantum
problem than a classical one.  in particular, a variety of diffusion
based, Monte Carlo methods are available that can be used to compute the
ground state energy of general quantum-mechanical systems.  Described in
detail elsewhere [5], these methods are based on the observation that
the Schr\"odinger equation written in imaginary time is isomorphic to
the diffusion equation with a growth depletion term.  In one-dimension
the result is
 \vskip .2in
\begin{equation}
{\partial \Psi \over \partial \tau} = { \hbar^2 \over \ 2m} {\partial^2 \Psi \over \partial x^2}-(V(x) - E_0)\Psi
\end{equation}
\vskip .2in
\noindent

where $\tau = it\hbar, E_0$ is a constant subtracted out for convenience
and $V(x)$ is the potential energy.

The Diffusion monte carlo (DMC) method is one, relatively simple
technique for treating such problems.  It follows the evolution of a
number of random walkers designed to move in such a way to simulate the
diffusion, growth and decay processes in Eq. (1).  Walkers in regions
where $E_0 < V(x)$ are attenuated via first-order decay, while those
in regions where $E_0 > V(x)$ undergo analogous growth.  In typical
applications, $E_0$ is adjusted iteratively to maintain a steady state
population.  In terms of the Schr\"odinger eigenfunctions and energies,
\{$\phi_n$\}, and \{$E_n$\}, respectively, the solution we seek is of
the form
\vskip .2in
\begin{equation}
\Psi(x, \tau) = \sum_n\phi_n(x)e^{-(E_n-E_0)\tau}
\end{equation}
\vskip .2in
\noindent
The ground state wavefunction can thus be identified as the
large $\tau$ limit of $\Psi(x, \tau)$ while the ground state energy is
equal to $E_0$.  One advantage of using DMC is that no knowledge of
the wavefunction is required.  Instead, it can be obtained from
the final distribution of walkers in the DMC simulation.

From Eq. (2), it is apparent that the rate of convergence of
the DMC method is controlled by the gap between the ground and first
excited state energies ($\Delta E)$.  In fact, the time required for
the DMC method to converge to the ground state is $\hbar/\Delta E$.
This is the same time scale needed for tunneling between two interacting
potential minima.  This suggests that DMC finds the ground state
wavefunction through tunneling.  By gradually increasing the mass of the
walkers thereby constraining the wave function, we can follow the
ground state energy to its classical limit.
\vskip .2in
\noindent

{\bf III.  Numerical Examples:}

\vskip .2in

To illustrate quantum annealing, we first consider the
problem of finding the minimum of the one-dimensional function shown in
Figures 2 and 3.  In Figure 2, we observe the evolution of transient
distributions of DMC diffusers at a fixed mass to demonstrate how the
method utilizes tunneling to find the ground state wave function.
Initially, all the diffusers are placed in the left well at a ground state
energy equal to approximately half the height of the intervening barrier
(line a).  Early in the simulation, some diffusers tunnel through the
barrier to the deeper, right well (line b).  Later, the population of
diffusers in the deeper well grows relative to the population in the left
well (line c) until the energy converges to the ground state and an
equilibrium distribution is reached (line d).  When the ground state
energy is large compared to the difference bewteen the two wells, the
distribution is almost symmetric.  However since the right well is
marginally deeper, the right population will be correspondingly larger
(assuming the number of diffusers is high enough to resolve the slight
difference).

In Figure 3, we show the evolution of the equilibrium
population of our model system as we turn off its quantum mechanical character
(lines a-c).  For a small mass, quantum effects are large and the
ground stae energy is relatively high compared with the slight energy
difference between the two wells.  In this case, the populations in
both wells are essentially equal (line a).  As we increase the mass, the
quantum character is reduced, the associated ground state energy
drops below the energy of the metastable minimum and the probability of
being found anywhere other than near the global minimum quickly dwindles
(line c).  In actual practice, it is generally not necessary to
completely converge to the ground state probability distribution before
increasing the system mass.  Once even a single DMC diffuser reaches a
particular potential well, it has the ability to proliferate and,
hence, sample that region.  As with conventional simulated annealing,
however, it is necessary to experiment with annealing rates to be sure that
results converge to the global minimum independent of initial
configuration chosen.

In our second example, we consider the problem of
determining the structure of N atom homogeneous Lennard-Jones clusters.  The
particles are assumed to interact pairwise via a two parameter
interaction of the form $\varepsilon V(\xi)$, where $\xi$ is a dimensionless
interparticle separation distance, $r/\sigma$.  The parameters
$\varepsilon$ and $\sigma$ are the usual Lennard-Jones energy and length scale
variables.  Explicitly, the Lennard-Jones potential is,
\vskip .2in
\begin{equation}
V(\xi) = 4(\xi^{-12} - \xi^{-6})
\end{equation}
\vskip .2in
\noindent
In terms of our reduced parameters, the quantum mechanical
Hamiltonian for our system is,
\vskip .2in
\begin{equation}
{1\over \varepsilon} H = - {1\over 2}\;\;\eta^2 \sum_{n=1}^N \nabla ^2_\xi + \sum_{m<n=1} V(\xi_{m,n})
\end{equation}
\vskip .2in
\noindent
where $\eta$ is a dimensionless constant that controls the
scale of the quantum mechanical effects,
\vskip .2in
\begin{equation}
\eta = {\hbar\over \sigma(m\varepsilon)^{1/2}}
\end{equation}
\vskip .2in
\noindent
For rare gases, $\eta$ ranges in magnitude from 0.425 for
$^4$He to 0.00995 for Xe.  As $\eta \rightarrow 0$, the ground state
energy approaches its classical limit.

Figure 4 shows that the calculated ground state energies for
Lennard-Jones clusters are relatively simple functions of
the parameter $\eta$.  It is not difficult to relate the limiting slopes
and curvatures of the ground state energies in Fig. (4) to
corresponding zero point energies and anharmonic effects, respectively.
The simplicity of the ground state energies is in marked
contrast to the excited states where the $\eta$-dependence is relatively
complex even for a 3 atom cluster [7].  We note in particular that the
ground state energy of the clusters varies smoothly with $\eta$ through
regions where the behavior of the clusters is changing from delocalized to
localized in character.  Since $\eta$ appears in Eq. (4) as the
coefficient of the highest order derivative, it is a ``singular perturbation,''
an indication that power series expansions in the parameter
$\eta$ may have a limited (possibly zero) radius of convergence.

The results of the quantum annealing calculations for
various cluster sizes are shown in Table I.  The energies quoted were
obtained by extrapolating a linear fit to the ground state energies for
small ($<$ 0.01) values of $\eta$.  The extrapolated potential minima
found are in good agreement with the known minimum energy structures of
the classical system [8].  We note that for all cases reported in Table I,
the agreement becomes exact if we refine the geometries of the
large mass DMC clusters using traditional methods, a relatively simple
task once the quantum annealing methods have successfully located the
vicinity of the global minimum.  That we obtain the correct structures is
non-trivial since the 13 atom cluster has 988 known minima, and the 19
atom cluster has on the order of 10$^5$ stable isomers [9].  We note that
the minimum energies found using the above extrapolation method are
upper bounds to the actual minima since we are using linear fits to
approximate a curve that is concave downward (see Figure 4).  Decreasing the
interval used in the fits ($\eta_{max}$) leads to a better extrapolation
of the curve and hence a better estimate of the potential minima.  For
example, in the seven particle cluster, as $\eta_{max}$ is decreased
from 0.05 to 0.02 to 0.01, the estimate for the potential minimum for the
goes from -16.469 to -16.495 to -16.505, respectively, the final value
being that reported by Hoare and Pal [8].
\vskip .2in
\noindent

{\bf IV.  Discussion:}

\vskip .2in
Any method of locating the global minimum must address the
issue of local minima.  Simulated annealing confronts this problem
through the device of classical ``thermal fluctuations.''  Quantum
annealing and other methods [3,4] use delocalization and tunneling to
avoid metastable regions.  By utilizing a quantum rather than a classical
system, the present approach exploits a number of specialized ground
state methods that are not available within classical problems.  Quantum
annealing has the further advantage of making knowledge of the
wavefunction unnecessary.  The physically different ways in which quantum
and simulated annealing avoid local minima suggests that these
type of approaches may complement each other in general optimization
applications.
\vskip .2in

{\bf Acknowledgements:}

\vskip .2in

This material is based upon work supported under two
National Science Foundation Graduate Research Fellowships, National Science
Foundation grant CHE-9203498, and the Petroleum Research Fund of the
American Chemical Society.

\newpage

%\documentstyle[12pt, myjcp]{report}
%\begin{document}
%\setlength{\topmargin}[-7mm}
\newcommand{\hangpar}{
\hangindent=0.3in
\hangafter=1
\noindent}
%\setlength{textheight}{9.5in}
%\setlength{\baselineskip}{24pt}
%\begin{document}
%\setlength{\mathindent}{0 in}
%\begin{document}
\end{doublespace}
\noindent
REFERENCES \\
\setlength{\parindent}{-0.5in}
\hangpar
[1]  W. H. Press, S. A. Teukolsky, W. T. Vetterling, and B.
P. Flannery, {\it Numerical Recipes in FORTRAN}, second ed.
(Cambridge University Press, Cambridge, 1992). \\
\hangpar
[2]  K. S. Kirkpatrick, C. D. Gelatt, and M. P. Vecchi,
Science {\bf 220} (1983) 671; K. S. Kirkpatrick, Journal of
Statistical Physics, {\ bf 34}, (1984) 975. \\
\hangpar
[3]  L. Piela, J. Kostrowicki, and H. Scheraga, J. Phys.
Chem. {\bf 93} (1989) 3339; R. J. Somorjai, J. Phys. Chem.
{\bf 95} (1991) 4141; K.
 A. Olszewski, L. Piela, and H. A. Scheraga, J. Phys. Chem.
{\bf 96} (1992) 4672. \\
\hangpar
[4]  P. Amara, D. Hsu, and J. E. Straub, J. Phys. Chem. {\bf
97} (1993) 6715.\\
\hangpar
[5]  D. Ceperley and B. Alder, Science {\bf 231} (1986) 555. \\
\hangpar
[6]  D. D. Frantz, D. L. Freeman, and J. D. Doll, J. Chem. Phys. {\bf
93} (1990) 2769. \\
\hangpar
[7]  D. L. Leiter, J. D. Doll, and R. M. Whitnell, J. Chem. Phys. {\bf
96} (1992) 9239. \\
\hangpar
[8]  M. R. Hoare and P. Pal, Adv. Phys. {\bf 20} (1971) 161. \\
\hangpar
[9] F. H. Stillinger and D. Stillinger, J. Chem. Phys. {\bf 93} (1990)
6013.
\renewcommand{\baselinestretch}{1.8in}
\doublespacing
%\end{document}

\newpage

\input figure.tex

\newpage

%\documentstyle[12pt, myjcp]{report}
%\begin{document}
\noindent
{\bf  Table I.  Minimum Potential Energy}
\vskip .2in
\noindent
This table shows the minimum energy in units of
$\varepsilon$ for
various Lennard-Jones clusters obtained via quantum
annealing and from
Hoare and Pal (see reference 5).  When conjugate gradient
methods are
used to refine the minimum, the Hoare and Pal results are
reproduced.
\vskip .2in
\begin{center}
\begin{tabular}{lcc}
\hline \hline
Number of particles&Quantum Annealing&Hoare and Pal \\
\hline
2&-0.998&-1.000 \\
3&-2.999&-3.000 \\
4&-5.997&-6.000 \\
5&-9.095&-9.104 \\
6&-12.710&-12.712 \\
7&-16.505&-16.505 \\
8&-19.794&-19.822 \\
9&-24.108&-24.113 \\
10&-28.408&-28.420 \\
13&-44.305&-44.327 \\
19&-72.622&-72.659 \\
\hline
\end{tabular}
\end{center}
%\end{document}

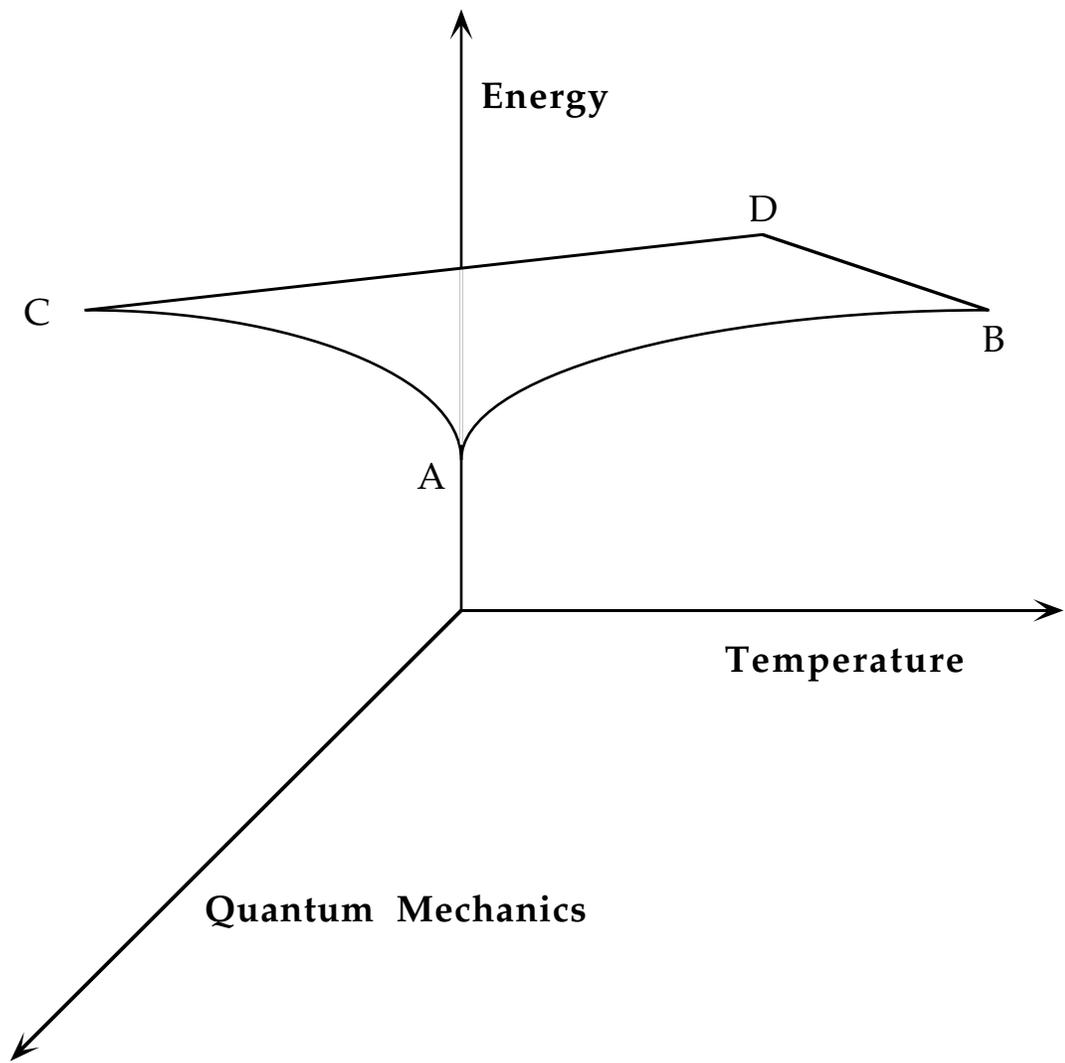
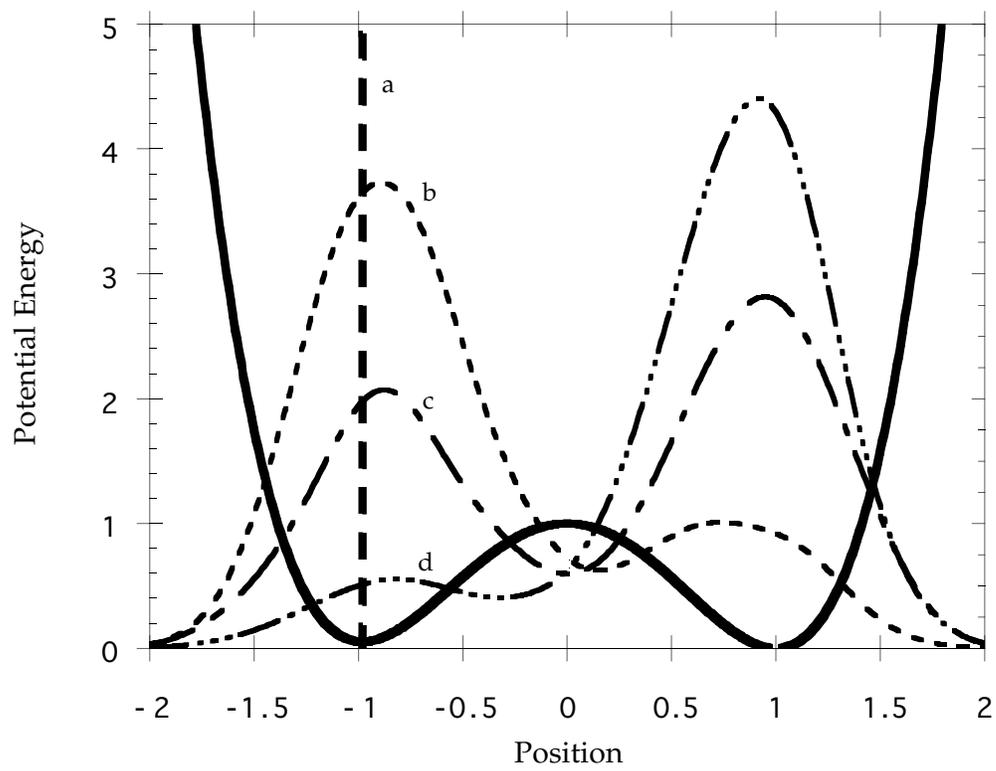
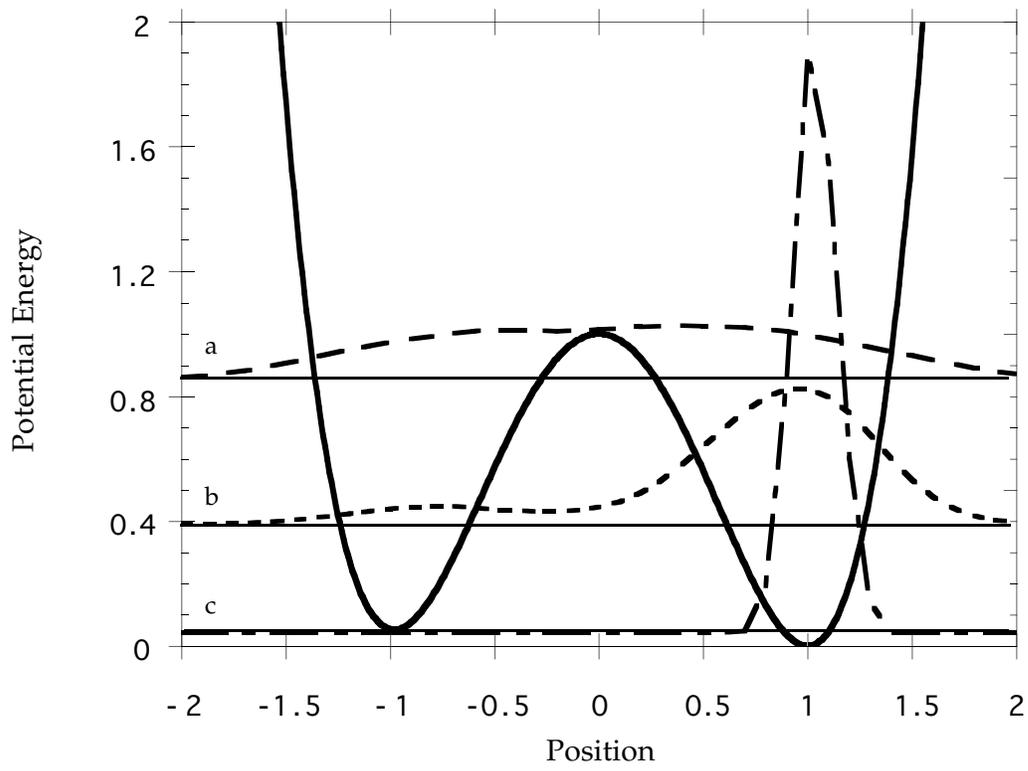
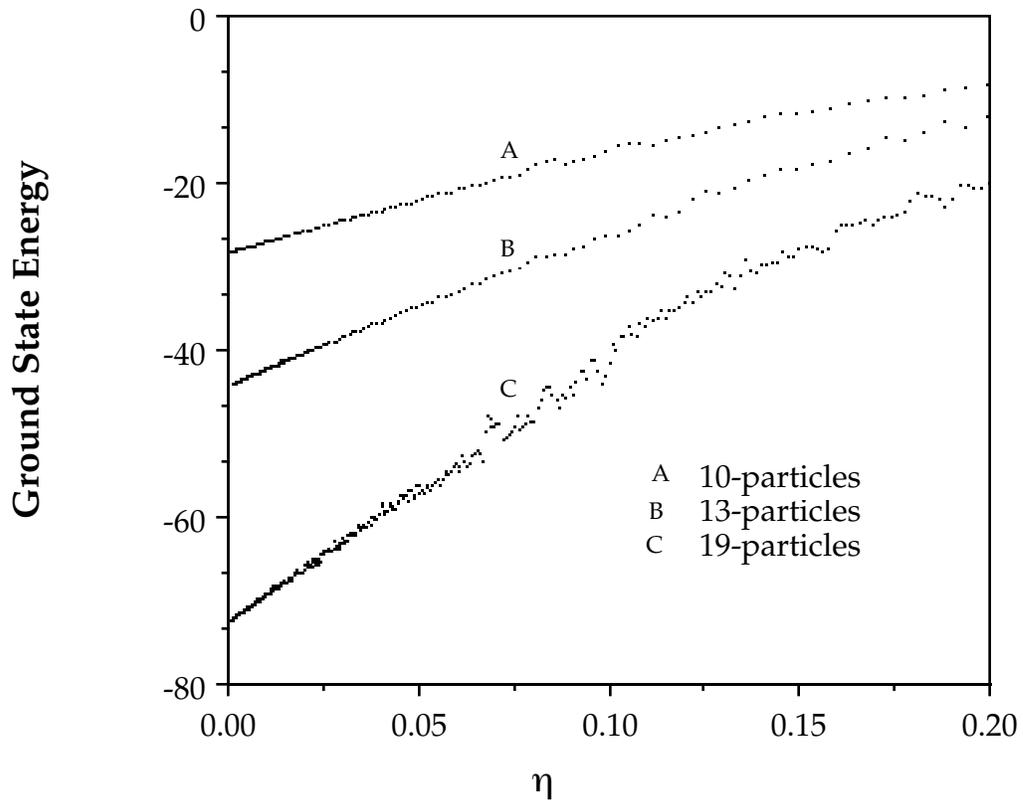
\end{document}